\documentclass[pra,showkeys,showpacs]{revtex4}
 \usepackage{graphics}
 \usepackage{graphicx}
\usepackage{amssymb}
\usepackage{latexsym}
\newtheorem{Theorem}{Theorem}[section]

\newtheorem{Lemma}[Theorem]{Lemma}

\newtheorem{Definition}[Theorem]{Definition}

\newcommand{\uline}{\vrule height.06ex depth.02ex width.6em}
\newcommand{\mvee}{\vee\kern-.69em\uline}

\begin{document}

\title{Mayet-Godowski Hilbert Lattice Equations}
\author{Norman D. Megill}
\email{nm@alum.mit.edu}
\homepage{http://www.metamath.org}
\affiliation{Boston Information Group, 19 Locke Ln., Lexington,
MA 02420, USA}
\author{Mladen Pavi\v ci\'c}
\email{pavicic@grad.hr}
\homepage{http://m3k.grad.hr/pavicic}
\affiliation{Physics Chair, Faculty of Civil Engineering,
University of Zagreb, Croatia}

\begin{abstract}
Several new results in the field of Hilbert lattice
equations based on states defined on the lattice as well as
novel techniques used to arrive at these results are presented.
An open problem of Mayet concerning Hilbert lattice
equations based on Hilbert-space-valued states is answered.
\end{abstract}


\maketitle

\section{Introduction}
\label{sec:intro}

When we deal with quantum systems and its theoretical
Hilbert space, we know that there is
a Hilbert lattice that is isomorphic to the set of
subspaces of any infinite-dimensional Hilbert
space and that we can establish a correspondence
between elements of the lattice and solutions of
a Schr\"odinger equation that corresponds to such a
Hilbert space. Therefore we might attempt to arrive
at an algebra which would enable us to introduce
quantum problems in a would-be quantum computer
in the same way in which we can introduce Boolean
algebra problem into a classical computer.
The gain is exponential---{\em any} quantum problem
could be solved in a polynomial time.

But there is an essential problem here.
Any Hilbert lattice is a structure based on
first-order predicate calculus,
and we simply cannot have a constructive pro\-ced\-ure
to introduce state\-ments like {\em there is} or {\em for all}
into a quantum computer. What we might do instead is to find
classes of polynomial lattice equations that can serve in
place of quantified statements. How far we have advanced down
this road we recently reviewed in Refs.~\cite{pavicic-book-05}
and \cite{pm-ql-l-hql2}, and in this paper we consider
recent results we obtained for particular classes of
such equations---Mayet-Godowski ones.

In 1985, Ren\'e Mayet \cite{mayet85} described a new equational variety
of lattices, which he called $OM_S^*$, that included all Hilbert
lattices and were included in a related variety of equations found by
Radoslaw Godowski \cite{godow} in 1981.  However, it was not known
whether the new variety was smaller than Godowski's (i.e., whether its
equations were independent from Godowski's).

Recently, the authors showed \cite{pm-ql-l-hql2} that Mayet's
variety is indeed strictly included in Godowski's.  In order to achieve
this result, several new algorithms had to be developed to find
counterexamples efficiently, to generate new equations in the family
that were violated by the counterexamples, and to prove that the new
equations were independent from every equation in the infinite family
found by Godowski.  This paper describes these algorithms, which were
incorporated into the computer programs that found this result.

In the last section, we show the solution to an open problem posed by
Mayet \cite{mayet06}, related to another new family of equations that he
found, based on strong sets of Hilbert-space-valued states.

\section{Definitions for lattice structures}
\label{sec:def1}

We briefly recall the definitions we will need.  For further
information, see Refs.~\cite{beran,mpoa99,pm-ql-l-hql1,pm-ql-l-hql2}.

\begin{Definition}\label{def:lattice}{\rm \cite{birk2nd}}
A {\em lattice} is an algebra
${\rm L}=\langle\mathcal{L}_{\rm O},\cap,\cup\rangle$
such that the following conditions are satisfied for any
$a,b,c\in\mathcal{L}_{\rm O}$:
\begin{eqnarray}
\begin{array}{ccc}
a\cup b=b\cup a\hfill&\hbox to 2cm{\hfill} & a\cap b=b\cap a\hfill\\
(a\cup b)\cup c=a\cup(b\cup c)
&&(a\cap b)\cap c=a\cap(b\cap c) \\
a\cap (a\cup b)=a\hfill &&a\cup (a\cap b)=a\hfill
\end{array}\nonumber
\end{eqnarray}
\end{Definition}

\begin{Theorem}\label{th:ordering}{\rm \cite{birk2nd}}
Binary relation $\le$ defined on {\rm L} as
\begin{eqnarray}
a\le b\quad {\buildrel{\rm def}\over\Longleftrightarrow}
\quad a=a\cap b\qquad\qquad
{\rm or\ as}\qquad\qquad
a\le b\quad {\buildrel{\rm def}\over\Longleftrightarrow}
\quad b=a\cup b
\end{eqnarray}
is a partial ordering
\end{Theorem}

\begin{Definition}{\rm \cite{birk3rd}}
An ortholattice {\rm (OL)} is an algebra
$\langle\mathcal{L}_{\rm O},',\cap,\cup,0,1\rangle$
such that $\langle\mathcal{L}_{\rm O},\cap,\cup\rangle$ is a lattice
with unary operation $'$ called {\em orthocomlementation}
which satisfies the following
conditions for $a,b\in\mathcal{L}_{\rm O}$ ($a'$ is called
the {\em orthocomplement} of $a$):
\begin{eqnarray}
&&a\cup a'=1, \qquad \qquad
a\cap a'=0\qquad \\
&&a\le b\quad \Rightarrow \quad b'\le a'\\
&& a''=a
\end{eqnarray}
\end{Definition}

\begin{Definition}\label{def:oml-o}{\rm \cite{pav93,p98}}
An {\em orthomodular lattice} {\em (OML)} is an {\em OL}
in which the following condition holds:
\begin{eqnarray}
a\leftrightarrow b=1\qquad\Leftrightarrow\qquad a=b
\label{eq:oml-le}
\end{eqnarray}
where  $a\leftrightarrow b=1 \ {\buildrel{\rm def}\over\Longleftrightarrow}
\ a\to b=1 \ \&\ \ b\to a=1$, where
$a\to b\ {\buildrel\rm def\over =}\ a'\cup(a\cap b)$
\end{Definition}

\begin{Definition}\label{def:commut}{\rm \cite{zeman}} We say
that $a$ and $b$ {\em commute} in {\em OML}, and write $aCb$,
when either of the following equivalent equations hold:
\begin{eqnarray}
a=((a\cap b)\cup (a\cap b'))\label{eq:commut1}\\
a\cap(a'\cup b)\le b\label{eq:commut2}
\end{eqnarray}
\end{Definition}

\begin{Definition}\label{def:hl}{\rm \protect{\footnote{For additional
definitions of the terms used in this section see
Refs.~\cite{beltr-cass-book,holl95,kalmb86,mpoa99}.}}}
An orthomodular lattice which satisfies the following
con\-di\-tions is a {\em Hilbert lattice}, $\mathcal{HL}$.
\begin{enumerate}
\item {\em Completeness:\/}
The meet and join of any subset of
an $\mathcal{HL}$ exist.
\item {\em Atomicity:\/}
Every non-zero element in an $\mathcal HL$ is greater
than or equal to an atom. (An atom $a$ is a non-zero lattice element
with $0< b\le a$ only if $b=a$.)
\item {\em Superposition principle:\/}
(The atom $c$
is a superposition of the atoms $a$ and $b$ if
$c\ne a$, $c\ne b$, and $c\le a\cup b$.)
\begin{description}
\item[{\rm (a)}] Given two different atoms $a$ and $b$, there is at least
one other atom $c$, $c\ne a$ and $c\ne b$, that is a superposition
of $a$ and $b$.
\item[{\rm (b)}] If the atom $c$ is a superposition of  distinct atoms
$a$ and $b$, then atom $a$ is a superposition of atoms $b$ and $c$.
\end{description}
\item {\em Minimal length:\/} The lattice contains at least
three elements $a,b,c$ satisfying: $0<a<b<c<1$.
\end{enumerate}
\end{Definition}

\begin{Definition}\label{def:state} A {\em state} {\rm (}also called
{\em probability measures} or simply {\em probabilities}
{\rm \cite{kalmb83,kalmb86,kalmb98,maczin}}{\rm )}
on a lattice $\mathcal L$
 is a function $m:{\mathcal L}\longrightarrow [0,1]$
such that $m(1)=1$ and $a\perp b\ \Rightarrow\ m(a\cup b)=m(a)+m(b)$,
where $a\perp b$ means $a\le b'$.
\end{Definition}

\begin{Lemma}\label{lem:state}
The following properties hold for any state $m$:
\begin{eqnarray}
&m(a)+m(a')=1\label{eq:state3}\\
&a\le b\ \Rightarrow\ m(a)\le m(b)\label{eq:state4}\\
&0\le m(a)\le 1\label{eq:state5}\\
&m(a_1)=\cdots=m(a_n)=1\
\Leftrightarrow\ m(a_1)+\cdots+m(a_n)=n\label{eq:state6}\\
&m(a_1\cap\cdots\cap a_n)=1\ \Rightarrow\
m(a_1)=\cdots=m(a_n)=1\label{eq:state7}
\end{eqnarray}
\end{Lemma}

\begin{Definition}\label{def:strong} A nonempty set $S$ of
states on $\mathcal L$ is called a strong set of states if
\begin{eqnarray}
(\forall a,b\in{\rm L})(\exists m \in S)((m(a)=1\ \Rightarrow
\ m(b)=1)\ \Rightarrow\ a\le b)\,.\quad
\label{eq:st-qm}
\end{eqnarray}
\end{Definition}

\begin{Theorem}\label{th:strong}{\rm \cite{mpoa99}} Every Hilbert
lattice admits a
strong set of states.
\end{Theorem}

\section{Definitions of equational families related to states}
\label{sec:def2}

First we will define the family of equations found by Godowski,
introducing a special notation for them.  These equations
hold in any lattice admitting a strong set of states
and thus, in particular, any Hilbert lattice. \cite{mpoa99}

\begin{Definition}\label{def:god-equiv}Let us call the
following expression the {\em Godowski identity}:
\begin{eqnarray}
a_1{\buildrel\gamma\over\equiv}a_n{\buildrel{\rm def}
\over =}(a_1\to a_2)\cap(a_2\to a_3)\cap\cdots
\cap(a_{n-1}\to a_n)\cap(a_n\to a_1),
\qquad
n=3,4,\dots\label{eq:god-equiv}
\end{eqnarray}
\end{Definition}

\begin{Theorem}\label{th:god-eq} Godowski's equations {\em\cite{godow}}
\begin{eqnarray}
a_1{\buildrel\gamma\over\equiv}a_3
&=&a_3{\buildrel\gamma\over\equiv}a_1
\label{eq:godow3o}\\
a_1{\buildrel\gamma\over\equiv}a_4
&=&a_4{\buildrel\gamma\over\equiv}a_1
\label{eq:godow4o}\\
a_1{\buildrel\gamma\over\equiv}a_5
&=&a_5{\buildrel\gamma\over\equiv}a_1
\label{eq:godow5o}\\
&\dots &\nonumber
\end{eqnarray}
hold in all ortholattices, {\em OL}'s, with strong sets of states.
An {\em OL} to which these equations are added is a variety
smaller than {\em OML}.
\end{Theorem}

We shall call these equations {\rm $n$-Go} {\rm (}{\rm 3-Go},
{\rm 4-Go}, etc.\/{\rm )}.  We also denote by
{\rm $n$GO} {\rm (}{\rm 3GO}, {\rm 4GO}, etc.\/{\rm )} the
{\rm OL} variety determined by {\rm $n$-Go} and
call it the {\rm $n$GO law}.

Next, we define a generalization of this family, first described by
Mayet. \cite{mayet85}  These equations also hold in all lattices
admitting a strong set of states, and in particular in all HLs.

\begin{Definition}\label{def:gge}
A {\em Mayet-Godowski equation} ({\rm MGE}) is an equality
with $n\ge 2$ conjuncts on each side:
\begin{eqnarray}
t_1 \cap \cdots\cap t_n = u_1 \cap \cdots\cap u_n
\end{eqnarray}
where each conjunct $t_i$ (or $u_1$) is a term consisting of
either a variable or a disjunction of two or more distinct
variables:
\begin{eqnarray}
t_i = a_{i,1}\cup \cdots\cup a_{i,p_i}\qquad \mbox{i.e. $p_i$ disjuncts}\\
u_i = b_{i,1}\cup \cdots\cup b_{i,q_i}\qquad \mbox{i.e. $q_i$ disjuncts}
\end{eqnarray}
and where the following conditions are imposed on the set of variables
in the equation:
\begin{enumerate}
\item{All variables in a given term $t_i$ or $u_i$ are
      mutually orthogonal.}
\item{ Each variable occurs the same number of times on each side of
       the equality.}
\end{enumerate}
\end{Definition}

We will call a lattice in which all MGEs hold an MGO; i.e., MGO is the
class (equational variety) of all lattices in which all MGEs hold.

The following three theorems about MGEs and MGOs are proved in
Ref.~\cite{pm-ql-l-hql2}.

\begin{Theorem}\label{th:mge}
A Mayet-Godowski equation holds in any ortholattice
 $\mathcal L$
ad\-mit\-ting a strong set of states and thus, in particular,
in any Hilbert lattice.
\end{Theorem}

\begin{Theorem}\label{th:mgo-in-ngo}
The family of all Mayet-Godowski equations
includes, in particular, the Godowski equations
{\rm [Eqs.~(\ref{eq:godow3o}),
(\ref{eq:godow4o}),\ldots]}; in other words, the class {\rm MGO}
is included in $n${\rm GO} for all $n$.
\end{Theorem}

\begin{Theorem}\label{th:mgo-lt-ngo}
The class {\rm MGO} is properly included in all $n${\rm GO}s, i.e.,
not all {\rm MGE} equations can be deduced from the equations
$n$-{\rm Go}.
\end{Theorem}

\begin{Definition}\label{def:stateeqn}
A {\em condensed state equation} is an abbreviated version of an MGE
constructed as follows:  all (orthogonality) hypotheses are discarded,
all meet symbols, $\cap$, are changed to $+$, and all join
 symbols, $\cup$, are changed to juxtaposition.
\end{Definition}

For example, the 3-Go equation can be expressed as: \cite{pm-ql-l-hql2}
\begin{eqnarray}
\lefteqn{a\perp d\perp b\perp e\perp c\perp f\perp a
    \qquad\Rightarrow} & & \nonumber \\
& & (a\cup d)\cap (b\cup e)\cap(c\cup f)\ =\
(d\cup b)\cap (e\cup c)\cap(f\cup a),
\label{eq:3gob}
\end{eqnarray}
which, in turn, can be expressed by the condensed state equation
\begin{eqnarray}
ad+be+cf&=&db+ec+fa.\label{eq:3goc}
\end{eqnarray}
The one-to-one correspondence between these
two representations of an MGE should be obvious.

\section{Checking $n$-Go equations on finite lattices}
\label{sec:dynamic}

For the general-purpose checking of whether an equation holds
in a finite lattice, the authors have primarily used a specialized
program, {\tt latticeg.c}, that is specialized to check an
equation provided by the user
against a list of Greechie diagrams (OMLs) provided by the user.  This
program has been described in Ref.~\cite{bdm-ndm-mp-1}.  While it has
proved essential to our work, a drawback is that the run time increases
quickly with the number of variables in and size of the input equation,
making it impractical for huge equations.

But there is another limitation in principle, not just in practice, for
the use of the {\tt latticeg.c} program.  In our work with MGEs, we are
particularly interested in those lattices having no strong set of states
but on which all of the successively stronger $n$-Gos pass, for all $n$
less than infinity.  This would prove that any MGE failing in that
lattice is independent from all $n$-Gos and thus represents a new
result.  The {\tt latticeg.c} program can, of course, check only a
finite number of such equations, and when $n$ becomes large the program
is too slow to be practical.  And in any case, it cannot provide a
proof, but only evidence, that a particular lattice does not violate
$n$-Go for any $n$.

Both of these limitations are overcome by a remarkable algorithm based
on dynamic programming, that was suggested by Brendan McKay.  This
algorithm was incorporated into a program, {\tt latticego.c}, that is
run against a set of lattices.  No equation is given to the program;
instead, the program tells the user the first $n$ for which $n$-Go fails
or whether it passes for all $n$.  The program runs very
quickly, depending only on the size of the input lattice, with a run
time proportional to the fourth power of the lattice size, rather than
increasing exponentially with $n$ as with the {\tt latticeg.c}
that checks against arbitrary equations.

To illustrate the algorithm, we will consider the specific case
of 7-Go.  From this example, the algorithm for the general case
of $n$-Go will be apparent.
We consider $7$-Go written in the following equivalent
 form: \cite{mpoa99}
\begin{eqnarray}
&& (a_1 \to a_2)\cap
 (a_2 \to a_3)\cap
 (a_3 \to a_4)\cap
 (a_4 \to a_5)\cap \nonumber\\
&& \qquad (a_5 \to a_6)\cap
 (a_6 \to a_7)\cap
 (a_7 \to a_1) \ \le\ a_1\to a_7 \label{eq:7go}
\end{eqnarray}
We define intermediate ``operations'' $E_1,\ldots,E_6$ along with
a predicate which provides the the answer:
\begin{eqnarray}
&&   E_1(a_1,a_2) = a_1\to a_2  \nonumber\\
&&   E_2(a_1,a_2,a_3) = E_1(a_1,a_2) \cap  (a_2\to a_3)   \nonumber\\
&&   E_3(a_1,a_2,a_3,a_4) = E_2(a_1,a_2,a_3) \cap  (a_3\to a_4)   \nonumber\\
&&   E_4(a_1,a_2,a_3,a_4,a_5) = E_3(a_1,a_2,a_3,a_4) \cap  (a_4\to a_5)
          \nonumber\\
&&   E_5(a_1,a_2,a_3,a_4,a_5,a_6) = E_4(a_1,a_2,a_3,a_4,a_5) \cap
           (a_5\to a_6)  \nonumber\\
&&   E_6(a_1,a_2,a_3,a_4,a_5,a_6,a_7) = E_5(a_1,a_2,a_3,a_4,a_5,a_6) \cap
            (a_6\to a_7)  \nonumber\\
&&   \mbox{answer}(a_1,a_7) = (E_6(a_1,a_2,a_3,a_4,a_5,a_6,a_7)
         \cap  (a_7\to a_1))
           \ \le\ (a_1\to a_7) \nonumber
\end{eqnarray}
Sets of values $V_2,\ldots,V_6$ are computed as follows:
\begin{eqnarray}
&&   V_2(a_1,a_3) = \{E_2(a_1,a_2,a_3) | a_2\}  \nonumber\\
&&   V_3(a_1,a_4) = \{E_3(a_1,a_2,a_3,a_4) | a_2,a_3\}  \nonumber\\
&&   V_4(a_1,a_5) = \{E_4(a_1,a_2,a_3,a_4,a_5) | a_2,a_3,a_4\}  \nonumber\\
&&   V_5(a_1,a_6) = \{E_5(a_1,a_2,a_3,a_4,a_5,a_6) | a_2,a_3,a_4,a_5\}
                 \nonumber\\
&&   V_6(a_1,a_7) = \{E_6(a_1,a_2,a_3,a_4,a_5,a_6,a_7) | a_2,a_3,a_4,a_5,a_6\}
             \nonumber\\
&&   \mbox{For all\ } a_1,a_7:  \mbox{answer}(a_1,a_7) \mbox{\ follows from\ }
              V_6(a_1,a_7), a_7\to a_1, \nonumber\\
&&   \qquad\qquad\mbox{\ and\ } a_1\to a_7   \nonumber
\end{eqnarray}
For example, $V_4(a_1,a_5)$ is the set of values
$E_4(a_1,a_2,a_3,a_4,a_5)$ can have
when $a_2,a_3,a_4$ range over all possibilities.  If
$\mbox{answer}(a_1,a_7)$ is true for all possible $a_1$ and $a_7$,
then 7-Go holds in the lattice, otherwise it fails.

The computation time is estimated as follows, where $n$ is the number
of nodes in the test lattice:
\begin{eqnarray}
&&   \mbox{Each\ } V_2(a_1,a_3)\mbox{\ can be found
       in\ }O(n)\mbox{\ time;\ }O(n^3)\mbox{\ total.} \nonumber\\
&&   \mbox{Each\ } V_3(a_1,a_4)\mbox{\ can be found
         in\ }O(n^2)\mbox{\ time from\ }V_2;
      O(n^4)\mbox{\ total.} \nonumber\\
&&   \mbox{Each\ } V_4(a_1,a_5)\mbox{\ can be found
         in\ }O(n^2)\mbox{\ time from\ }V_3;
      O(n^4)\mbox{\ total.}   \nonumber\\
&&  \qquad\qquad\qquad       \vdots   \nonumber
\end{eqnarray}
So the total time is $O(n^4)$.

The program is written so that it only has to compute additional ``inner
terms'' to process the next $n$-Go equation.  Remarkably, when a lattice
does not violate any $n$-Go, the addition of new terms tends to converge
to a fixed value rather quickly, meaning that $V_{n}$ for $(n+1)$-Go
remains the same as $V_{n-1}$ for $n$-Go.  This almost always happens
for $n<10$, and when it does, we can terminate the algorithm and say
with certainty that no further increase in $n$ will cause an $n$-Go
equation to fail in the lattice.  (If it doesn't happen, the program
will tell us that, but such a case has so far not been observed.  The
program has an arbitrary cut-off point of $n=100$, after which the
algorithm will terminate.  Observed runs have always either converged or
failed far below this point, and in any case the cut-off can be
increased with a parameter setting.)  Convergence provides a proof
that the entire class of Godowski equations (for {\em all} $n<\infty$)
will pass in the lattice.  Such a feat is not possible with
ordinary lattice-checking programs, since an infinite number of
equations would have to be tested.

When a lattice does violate some $n$-Go, that result tends to be found
even faster, and the algorithm terminates, and the program tells us the
first $n$ at which an $n$-Go equation fails in the lattice.  Since
$n$-Go can be derived from $n+1$-Go, failure is also implied for all
greater $n$.

The success of {\tt latticego.c} depends crucially on the structure of a
particular representation of the $n$-Go equations, where variables
appear only on one side of the equation and are localized to an adjacent
pair of conjuncts in a chain of conjuncts.  So far, efforts to adapt the
approach to other equational families, such as the $n$OA (generalized
orthoarguesian) laws, \cite{mpoa99} haven't been successful but are
still being explored.

\section{Finding states on finite lattices}
\label{sec:states}

It is possible to express the set of constraints imposed by states
as a   linear programming (LP) problem.  Linear programming is used by
industry to minimize cost, labor, etc., and many efficient programs have
been developed to solve these problems, most of them based on the
simplex algorithm.

We will examine a particular example in detail to illustrate how the
problem is expressed.  For this example we will consider a Greechie
diagram with 3-atom blocks, although the principle is easily extended
to any number of blocks.

If $m$ is a state, then each 3-atom block with atoms $a$, $b$, $c$ and
complements $a'$, $b'$, $c'$ imposes the following constraints:
\begin{eqnarray}
  m(a) + m(b) + m(c)&\ =\ &1 \label{eqn:mmm} \\
  m(a') + m(a)&\ =\ &1       \nonumber\\
  m(b') + m(b)&\ =\ &1       \nonumber\\
  m(c') + m(c)&\ =\ &1       \nonumber\\
  m(x) &\ \ge\ & 0,\qquad x=a,b,c,a',b',c' \nonumber
\end{eqnarray}
To obtain Eq.~(\ref{eqn:mmm}), note that in any Boolean block,
$a \perp b \perp c \perp a$, so $m(a) = 1 - m(a') =
1 - m(b\cup c) = 1 - m(b) - m(c)$.

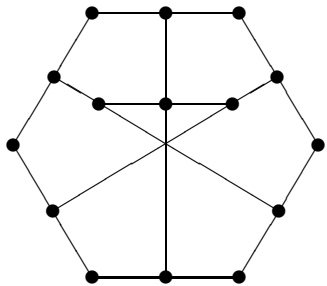
\begin{figure}[htbp]\centering
  \begin{picture}(250,115)(0,0)

   \put(0,0) { 
      \begin{picture}(124,110)(0,0) 
        \put(32.2,0){\line(1,0){55.6}}
        \put(32.2,100){\line(1,0){55.6}}
        \put(2.3,50){\line(3,5){29.9}}
        \put(117.7,50){\line(-3,5){29.9}}
        \put(2.3,50){\line(3,-5){29.9}}
        \put(117.7,50){\line(-3,-5){29.9}}
        \put(60,100){\line(0,-1){100}}
        \put(17.25,25){\line(5,3){84.8}}
        \put(102.75,25){\line(-5,3){84.8}}
        \put(34.71,65.6){\line(1,0){50.58}}

        \put(2.3,50){\circle*{5}}
        \put(117.7,50){\circle*{5}}
        \put(87.8,0){\circle*{5}}
        \put(87.8,100){\circle*{5}}
        \put(32.2,0){\circle*{5}}
        \put(32.2,100){\circle*{5}}
        \put(60,0){\circle*{5}}
        \put(60,100){\circle*{5}}
        \put(17.25,25){\circle*{5}}
        \put(17.85,76){\circle*{5}}
        \put(102.75,25){\circle*{5}}
        \put(102.15,76){\circle*{5}}
        \put(34.71,65.6){\circle*{5}}
        \put(85.29,65.6){\circle*{5}}
        \put(60,65.6){\circle*{5}}
      \end{picture}
    } 

  \end{picture}
  \caption{Greechie diagram for the Peterson lattice.\label{fig:peterson}}
\end{figure}

Let us take the specific example of the Peterson lattice, which we
know does not admit a set of strong states.  The Greechie diagram
for this lattice, shown in Fig.~\ref{fig:peterson},
can be expressed with the textual notation
\begin{verbatim}
    123,345,567,789,9AB,BC1,2E8,4FA,6DC,DEF.
\end{verbatim}
(see Ref.~\cite{pm-ql-l-hql2}), where each digit or letter represents
an atom, and groups of them represent blocks (edges of the Greechie
diagram).

Referring to the textual notation,
we designate the atoms by $1,2,\ldots,F$ and their orthocomplements by
$1',2',\ldots,F'$.  We will represent the values of state $m$ on the
atoms by $m(1), m(2), \ldots, m(F)$.  This gives us the following
constraints for the 10 blocks:
\begin{eqnarray}
  m(1) + m(2) + m(3)&\ =\ &1   \nonumber\\
  m(3) + m(4) + m(5)&\ =\ &1   \nonumber\\
  m(5) + m(6) + m(7)&\ =\ &1   \nonumber\\
  m(7) + m(8) + m(9)&\ =\ &1   \nonumber\\
  m(9) + m(A) + m(B)&\ =\ &1   \nonumber\\
  m(B) + m(C) + m(1)&\ =\ &1   \nonumber\\
  m(2) + m(E) + m(8)&\ =\ &1   \nonumber\\
  m(4) + m(F) + m(A)&\ =\ &1   \nonumber\\
  m(6) + m(D) + m(C)&\ =\ &1   \nonumber\\
  m(D) + m(E) + m(F)&\ =\ &1   \nonumber
\end{eqnarray}
In addition, we have $m(a') + m(a) = 1$, $m(a) \ge 0$, and $m(a')\ge 0$
for each atom $a$, adding potentially an additional $15\times 3=45$
constraints.  However, we can omit all but one of these since most
orthocomplemented atoms are not involved this problem, the given
constraints are sufficient to ensure that the state values for atoms are
less than 1, and the particular linear programming algorithm
we  used assumes all variables
are nonnegative.  This speeds up the computation considerably.  The only
one we will need is $m(7) + m(7')=1$ because, as we will see, the
orthocomplemented atom $7'$ will be part of the full problem statement.

We pick two incomparable nodes, 1 and $7'$, which are on opposite sides
of the Peterson lattice.  (The program will try all possible
pairs of incomparable nodes, but for this example we have selected
a priori a pair that
will provide us with the answer.)  Therefore it is the case that $ \sim
1 \le 7'$.  If the Peterson lattice admitted a strong set of states, for
any state $m$ we would have:
\begin{eqnarray}
  (m(1) = 1 \quad\Rightarrow\quad m(7') = 1)
      \quad\Rightarrow\quad 1 \le 7'. \nonumber
\end{eqnarray}

Since the the conclusion is false, for some $m$ we must have
\begin{eqnarray}
&&  \sim (m(1) = 1 \quad\Rightarrow\quad  m(7') = 1)      \nonumber\\
&&  {\rm i.e.}\ \ \sim (\sim\ m(1) = 1 \quad\vee\quad m(7') = 1)  \nonumber\\
&&  {\rm i.e.}\ \ \ m(1) = 1 \quad\&\quad \sim\ m(7') = 1         \nonumber
\end{eqnarray}
So this gives us another constraint:
\begin{eqnarray}
  m(1) = 1;       \nonumber
\end{eqnarray}
and for a set of strong states to exist there must be some m such that
\begin{eqnarray}
  m(7) < 1.     \nonumber
\end{eqnarray}
So, our final linear programming
problem becomes (expressed in the notation of the
publicly available program {\tt lp\_solve}\protect{\footnote{Available at
{\tt http://groups.yahoo.com/group/lp\_solve/files/
(Sep\-tember 2006).}}}):
\begin{verbatim}
  min: m7';
  m1 = 1;
  m7 + m7' = 1;
  m1 + m2 + m3 = 1;
  m3 + m4 + m5 = 1;
  m5 + m6 + m7 = 1;
  m7 + m8 + m9 = 1;
  m9 + mA + mB = 1;
  mB + mC + m1 = 1;
  m2 + mE + m8 = 1;
  m4 + mF + mA = 1;
  m6 + mD + mC = 1;
  mD + mE + mF = 1;
\end{verbatim}
which means ``minimize $m(7')$, subject to constraints $m(1) = 1,
m(7)+m(7')=1, \ldots$.''  The variable to be minimize, $m(7)$, is called
the {\em objective function} (or ``cost function'').  When this problem
is given to {\tt lp\_solve}, it returns an objective function value of
1. This means that regardless of $m$, the other constraints impose a
minimum value of 1 on $m(7')$, contradicting the requirement that $m(7')
< 1$.  Therefore, we have a proof that the Peterson lattice does not
admit a set of strong states.

The program {\tt states.c} that we use reads a list of Greechie diagrams
and, for each one, indicates whether or not it admits a strong set of
states.  The program embeds the {\tt lp\_solve} algorithm, wrapping
around it an interface that translates, internally, each Greechie
diagram into the corresponding linear programming problem.

\section{Generation of MGEs from finite lattices}
\label{sec:gen}

When the linear programming problem in the previous section finds a pair
of incomparable nodes that prove that the lattice admits no strong set
of states, the information in the problem can be used to find an
equation that holds in any OML admitting a strong set of states, and in
particular in HL, but fails in the OML under test.  Typically, an OML to
be tested was chosen because it does not violate any other known HL
equation.  Thus, by showing an HL equation that fails in the OML under
test, we will have found a new equation that holds in HL and is
independent from other known equations.

The set of constraints that lead to the objective function value of 1 in
our linear programming problem turns out to be redundant.  Our algorithm
will try to find a minimal set of hypotheses (constraints) that are
needed.  The equation-finding mode of {\tt states.c} program
incorporates this algorithm, which will try to weaken the constraints of
the linear programming problem one at a time, as long as the objective
function value remains 1 (as in the problem in the previous section).
The equation will be constructed based on a minimal set of unweakened
constraints that results.

The theoretical basis for the construction is described in the proof of
Theorem 30 of Ref.~\cite{pm-ql-l-hql2}.  Here, we will describe
the algorithm by working through a detailed example.

Continuing from the final linear programming problem of the previous
section, the program will test each constraint corresponding to a
Greechie diagram block, i.e.\ each equation with 3 terms, as follows.
It will change the right-hand side of the constraint equation from $= 1$
to $\le 1$, thus weakening it, then it will run the linear programming
algorithm again.  If the weakened constraint results in an objective
function value $m(7')<1$, it means that the constraint is needed to
prove that the lattice doesn't admit a strong set of states, so we
restore the r.h.s.\ of that constraint equation back to 1. On the other
hand, if the objective function value remains $m(7')=1$ (as in the
original problem), a tight constraint on that block is not needed for
the proof that the lattice doesn't admit a strong set of states, so we
leave the r.h.s.\ of that constraint equation at $\le 1$.

After the program completes this process, the linear programming problem
for this example will look like this:
\begin{verbatim}
min: m7';
m1 = 1;
m7 + m7' = 1;
m1 + m2 + m3 <= 1;
m3 + m4 + m5 = 1;
m5 + m6 + m7 <= 1;
m7 + m8 + m9 <= 1;
m9 + mA + mB = 1;
mB + mC + m1 <= 1;
m2 + mE + m8 = 1;
m4 + mF + mA <= 1;
m6 + mD + mC = 1;
mD + mE + mF <= 1;
\end{verbatim}
Six out of the 10 blocks have been made weaker, and the linear
programming algorithm will show that the objective function has remained
at 1. We now have enough information to construct the MGE, which we will
work with in the abbreviated form of a condensed state equation
(Definition~\ref{def:stateeqn}).
\begin{enumerate}
\item Since $m(1)=1$, the other atoms in the two blocks (3-term
    equations) using it will be $0$.  Thus $m(2)=m(3)=m(B)=m(C)=0$.
\item For each of the four blocks that have $=1$ on the r.h.s., we
    suppress the atoms that are $0$ and juxtapose the remaining 2 atoms in
    each block.  For example, in $m(3)+m(4)+m(5)=1$, we ignore $m(3)=0$, and
    collect the atoms from the remaining two terms to result in $45$ (4
    juxtaposed with 5).  Then we join all four pairs with $+$ to build the
    l.h.s.\ form for the condensed state equation:
    \begin{eqnarray}
      45+9A+E8+6D   \label{eq:lhs}
    \end{eqnarray}
\item For the r.h.s.\ of the equation, we scan the blocks with weakened
    constraints.  From each block, we pick out and juxtapose those
    atoms that also appear on the l.h.s.
    and discard the others.   For example, in $m(5)+m(6)+m(7)\le 1$,
    5 and 6 appear in Eq.~(\ref{eq:lhs}) but 7 doesn't.
    Joining the juxtaposed groups with $+$, we build the r.h.s.:
    \begin{eqnarray}
      56+89+4A+DE    \nonumber
    \end{eqnarray}
    Note that out of the 6 weakened constraints, 2 of them have no
    atoms at all in common with \ Eq.~(\ref{eq:lhs}) and are
    therefore ignored.
\item
    Equating the two sides, we obtain the form of the condensed
     state equation:
    \begin{eqnarray}
      45+9A+E8+6D=56+89+4A+DE  \nonumber
    \end{eqnarray}
\item
    Replacing the atoms with variables, the final condensed state
    equation becomes:
    \begin{eqnarray}
      ab+cd+ef+gh=bg+fc+ad+he  \label{eq:state4go}
    \end{eqnarray}
\item
  Finally, the number of occurrences of each variable on must match on
  each side of the condensed state equation.  In this example, that is
  already the case.  But in general, there may be terms that will have to
  be repeated in order to make the numbers balance.  An example with such
  ``degenerate'' terms is shown as Eq.~(47) of
  Ref.~\cite{pm-ql-l-hql2}.
\end{enumerate}
Eq.~(\ref{eq:state4go}) will be recognized, after converting it to
an MGE, as the 4-Go equation, which as is well-known holds in
all OMLs that admit a strong set of states but fails in the
Peterson lattice Fig.~\ref{fig:peterson}. \cite{mpoa99}

\section{Solution to an open problem}
\label{sec:open}

In Ref.~\cite{mayet06}, Mayet shows the following consequence of one of his
equations ($E^*_2$) derived from Hilbert-space-valued states, and asks
whether an OML exists in which it fails.  The answer is negative.
\begin{Theorem}\label{th:mayetprob}
The condition
\begin{eqnarray}
&& a_1\perp b_1\ \&\ a_2\perp b_2\ \&\ a_1\perp a_2 \nonumber\\
&& \qquad  \quad\Rightarrow\quad (a_1\cup b_1)\cap(a_2\cup b_2)
  \le b_1\cup b_2\cup (a_1\cup a_2)' \label{eq:mayetprob}
\end{eqnarray}
holds in all {\rm OML}s.
\end{Theorem}
{\em Proof.}
The following lemma follows using DeMorgan's law,
the Foulis-Holland theorem (F-H; see e.g.\ Ref.~\cite[p.~25]{kalmb83}),
and the orthogonality
hypothesis $a_2\perp b_2$, respectively, for its three steps.  The
orthogonality hypotheses provide the commute relations
needed for F-H.
\begin{eqnarray}
 b_2 \cup (a_1 \cup a_2)' &\ =\ & b_2 \cup (a_1' \cap a_2')\nonumber\\
                   &\ =\ & (b_2 \cup a_1')\cap (b_2 \cup a_2') \nonumber\\
                   &\ =\ & (b_2 \cup a_1')\cap a_2'. \label{eq:mayetproof1}
\end{eqnarray}
From $(b_1 \cap a_1')\cup (b_2 \cup a_1')= a_1' \cup b_2$
(in any OL) and
$(b_1 \cap a_1')\cup a_2' = a_2' \cup b_1$ (using $a_1\perp b_1$),
we have
\begin{eqnarray}
 (a_1' \cup b_2)\cap (a_2' \cup b_1)
  &\  =\ &((b_1 \cap a_1')\cup (b_2 \cup a_1'))
       \cap ((b_1 \cap a_1')\cup a_2').\label{eq:mayetproof2}
\end{eqnarray}
The result then follows by applying the hypotheses, OL,
Eq.~\ref{eq:mayetproof2}, F-H, the hypothesis $a_1\perp b_1$, and
Eq.~\ref{eq:mayetproof1} to obtain the following steps, respectively:
\begin{eqnarray}
 (a_1 \cup b_1)\cap (a_2 \cup b_2)
  &\  \le\ & (a_2' \cup b_1)\cap (a_1' \cup b_2)  \nonumber\\
  &\  =\ & (a_1' \cup b_2)\cap (a_2' \cup b_1)    \nonumber\\
  &\  =\ & ((b_1 \cap a_1')\cup (b_2 \cup a_1'))
            \cap ((b_1 \cap a_1')\cup a_2')  \nonumber\\
  &\  =\ &(b_1 \cap a_1')\cup ((b_2 \cup a_1')\cap a_2')
        \nonumber\\
  &\  =\ & b_1 \cup ((b_2 \cup a_1')\cap a_2')   \nonumber\\
  &\  =\ & b_1 \cup b_2 \cup (a_1 \cup a_2)'. \nonumber
\end{eqnarray}
\hfill$\Box$

\section{Conclusion}
\label{sec:concl}

In the previous sections we presented several results obtained in the
field of Hilbert space equations based on the states defined on the
space.  The idea is to use classes of Hilbert lattice equations for an
alternative representation of Hilbert lattices and Hilbert spaces of
arbitrary quan\-tum systems that might eventually enable a direct
introduction of the states of the systems into quantum computers.  In
applications, infinite classes could then be ``truncated'' to provide us
with finite classes of required length.  The obtained classes would in
turn contribute to the theory of Hilbert space subspaces, which so far
is poorly developed.

In Sections \ref{sec:def2}--\ref{sec:open} we have considered three
classes of Hilbert lattice equations based on the states defined on the
lattice by means of Definition \ref{def:state} and specified in Section
\ref{sec:open}.

The algorithms and associated computer programs that were developed for
this project were essential to its success.  McKay's dynamic programming
algorithm for $n$-Go equations (Section \ref{sec:dynamic}), together
with its quickly convergent behavior for large $n$, was particularly
fortuitous.  Without it, the authors see no apparent way that the
independence of the MGE equations from all $n$-Go equations could be
answered.  At best, only empirical evidence pointing towards that answer
could be accumulated, but that of course would not constitute a proof.
Indeed, this problem had remained open for nearly 20 years since Mayet's
first publication \cite{mayet85} of these equations.

Thus, there is a strong motivation to find variants of McKay's $n$-Go
dynamic programming algorithm that could be more generally applied to
other infinite families, in particular the $n$OA
(generalized orthoarguesian) laws.  \cite{mpoa99} Assuming
similar run-time behavior could be achieved, this would provide us with
an extremely powerful tool that would let us test finite lattices
against the family very quickly (instead of months or years of CPU time)
as well as prove independence results for the entire infinite family of
equations at once (if the valuation set rapidly converges to a final,
fixed value with increasing $n$, as it does for $n$-Go).  On the surface
this appears to be quite a difficult problem, because unlike the special
form in which the $n$-Go equations can be rewritten - so that most of
the variables are localized to adjacent conjuncts - the known forms of
the $n$OA laws have their variables distributed throughout their (very
long) equations.  So another approach would be to discover a new form of
the $n$OA laws that better separates their variable occurrences in such
a way that a variant of the $n$-Go dynamic programming algorithm might
be applicable.  Both of these approaches are being investigated by the
authors.

The authors are unaware of any previous use of linear programming
methods for finding states on a finite lattice and in particular (for
the present study) determining whether the lattice admits a strong set
of states.  There appear to be relatively few programs that deal with
states, and most of the finite lattice examples in the literature
related to states were found by hand.  A Pascal program written by Klaey
\cite{klaey} is able to find certain kinds of states on lattices, but
for the strong set of states problem it is apparently able only to indicate
``yes'' (if a strong set of states was found) or ``unknown''
otherwise.  The linear programming method provides a definite answer
either way, in the predictable amount of time that the simplex algorithm
takes to run.  Finally, the  linear programming problem
itself (with redundant constraints weakened) provides us with the
information we need to construct a new Hilbert lattice equation that
fails in a given lattice not admitting a strong set of states.

Mayet's recent and important $E$-equation results \cite{mayet06} provide
us with powerful new method, the use of Hilbert-space-valued states, to
find previously unknown families of equations that hold in Hilbert
lattices.  For further investigation of these equations, it will be
highly desirable to have a program analogous to our {\em states.c}
(which works with only real-valued states) that will tell us whether or
not a finite lattice admits a strong set of Hilbert-space-valued states.
This problem seems significantly harder than that of finding real-valued
states, and possible algorithms for doing this are being explored by the
authors.

The programs {\tt latticego.c} and {\tt states.c} described above can
be downloaded from {\tt http://us.metamath.org/\#downloads}.

\begin{acknowledgments}
Supported by the Ministry of
Science, Education, and Sport of Croatia.
\end{acknowledgments}

\bibliographystyle{report}

\begin{thebibliography}{10}

\bibitem{pavicic-book-05}
M.~Pavi{\v c}i{\'c},
\newblock {\em Quantum Computation and Quantum Communication: {T}heory and
  Experiments},
\newblock Springer, New York, 2005.

\bibitem{pm-ql-l-hql2}
M.~Pavi{\v c}i{\'c} and N.~D. Megill,
\newblock {\it Quantum Logic and Quantum Computation},
\newblock in {\em Handbook of Quantum Logic}, edited by K.~Engesser, D.~Gabbay,
  and D.~Lehmann, volume~2, pages 751--787, Elsevier, Amsterdam, 2006.

\bibitem{mayet85}
R.~Mayet,
\newblock Varieties of Orthomodular Lattices Related to States,
\newblock {\it Algebra Universalis} {\bf {\bf 20}}, 368--396 (1985).

\bibitem{godow}
R.~Godowski,
\newblock Varieties of Orthomodular Lattices with a Strongly Full Set of
  States,
\newblock {\it Demonstratio Math.} {\bf {\bf 14}}, 725--733 (1981).

\bibitem{mayet06}
R.~Mayet,
\newblock Equations Holding in {H}ilbert Lattices,
\newblock {\it Int. J. Theor. Phys.} {\bf {\bf 45}}, 1216--1246 (2006).

\bibitem{beran}
L.~Beran,
\newblock {\em Orthomodular Lattices; {A}lgebraic Approach},
\newblock D. Reidel, Dordrecht, 1985.

\bibitem{mpoa99}
N.~D. Megill and M.~Pavi{\v c}i{\'c},
\newblock Equations, States, and Lattices of Infinite-Dimensional {H}ilbert
  Space,
\newblock {\it Int. J. Theor. Phys.} {\bf {\bf 39}}, 2337--2379 (2000).

\bibitem{pm-ql-l-hql1}
M.~Pavi{\v c}i{\'c} and N.~D. Megill,
\newblock {{\it Is Quantum Logic a Logic?}},
\newblock in {\em Handbook of Quantum Logic}, edited by K.~Engesser, D.~Gabbay,
  and D.~Lehmann, volume~1, Elsevier, Amsterdam, 2006.

\bibitem{birk2nd}
G.~Birkhoff,
\newblock {\em Lattice Theory}, volume XXV of {\em American Mathematical
  Society Colloqium Publications},
\newblock American Mathematical Society, New York, 2nd (revised) edition, 1948.

\bibitem{birk3rd}
G.~Birkhoff,
\newblock {\em Lattice Theory}, volume XXV of {\em American Mathematical
  Society Colloquium Publications},
\newblock American Mathematical Society, Providence, Rhode Island, 3rd (new)
  edition, 1967.

\bibitem{pav93}
M.~Pavi{\v c}i{\'c},
\newblock Nonordered Quantum Logic and Its {YES}--{NO} Representation,
\newblock {\it Int. J. Theor. Phys.} {\bf {\bf 32}}, 1481--1505 (1993).

\bibitem{p98}
M.~Pavi{\v c}i{\'c},
\newblock Identity Rule for Classical and Quantum Theories,
\newblock {\it Int. J. Theor. Phys.} {\bf {\bf 37}}, 2099--2103 (1998).

\bibitem{zeman}
J.~J. Zeman,
\newblock Quantum Logic with Implications,
\newblock {\it Notre Dame J. Formal Logic} {\bf {\bf 20}}, 723--728 (1979).

\bibitem{kalmb83}
G.~Kalmbach,
\newblock {\em Orthomodular Lattices},
\newblock Academic Press, London, 1983.

\bibitem{kalmb86}
G.~Kalmbach,
\newblock {\em Measures and Hilbert Lattices},
\newblock World Scientific, Singapore, 1986.

\bibitem{kalmb98}
G.~Kalmbach,
\newblock {\em Quantum Measures and Spaces},
\newblock Kluwer, Dordrecht, 1998.

\bibitem{maczin}
M.~J. M{\c a}czy{\'n}ski,
\newblock Hilbert Space Formalism of Quantum Mechanics without the {H}ilbert
  Space Axiom,
\newblock {\it Rep. Math. Phys.} {\bf {\bf 3}}, 209--219 (1972).

\bibitem{bdm-ndm-mp-1}
B.~D. Mc{K}ay, N.~D. Megill, and M.~Pavi{\v c}i{\'c},
\newblock Algorithms for {G}reechie Diagrams,
\newblock {\it Int. J. Theor. Phys.} {\bf {\bf 39}}, 2381--2406 (2000).

\bibitem{klaey}
M.~Kl{\"a}y,
\newblock {\em Stochastic Models on Empirical Systems, Empirical Logic and
  Quantum Logics, and States on Hypergraphs},
\newblock PhD thesis, University of {B}ern, {F}aculty of {N}atural {S}cience,
  Fischer {D}ruck,{M}{\"u}nsingen, 1985.

\bibitem{beltr-cass-book}
E.~G. Beltrametti and G.~Cassinelli,
\newblock {\em The Logic of Quantum Mechanics},
\newblock Addison-Wesley, 1981.

\bibitem{holl95}
S.~S. {Holland, JR.},
\newblock Orthomodularity in Infinite Dimensions; a Theorem of {M}.
  {S}ol{\`e}r,
\newblock {\it Bull. Am. Math. Soc.} {\bf {\bf 32}}, 205--234 (1995).

\end{thebibliography}

\end{document}